\documentclass[12pt]{article}
\usepackage[utf8]{inputenc}
\usepackage{geometry}
\usepackage{lmodern}
\usepackage{amsmath, amssymb}
\usepackage{graphicx}
\usepackage{hyperref}
\title{Reducing Procrastination on Programming Assignments via Optional Early Feedback}
\author{
Alice Gao\\
\texttt{ax.gao@utoronto.ca}\\
University of Toronto\\
Toronto, Ontario, Canada\\[1em]
Victoria Sakhnini\\
\texttt{vsakhnini@uwaterloo.ca}\\
University of Waterloo\\
Waterloo, Ontario, Canada
}
\date{}
\begin{document}
\maketitle

\section*{Abstract}
Academic procrastination is prevalent among undergraduate computer science students. Many studies have linked procrastination
to poor academic performance and well-being. Procrastination is
especially detrimental for advanced students when facing large,
complex programming assignments in upper-year courses. We designed an intervention to combat academic procrastination on such
programming assignments. The intervention consisted of early
deadlines that were not worth marks but provided additional automated feedback if students submitted their work early. We evaluated
the intervention by comparing the behaviour and performance of
students between a control group and an intervention group. Our
results showed that the intervention encouraged significantly more
students to start the assignments early. Although there was no
significant difference in students' grades between the control and
intervention groups, students within the intervention group who
used the intervention achieved significantly higher grades than
those who did not. Our results implied that starting early alone did
not improve students' grades. However, starting early and receiving additional feedback enhanced the students' grades relative to
those of the rest of the students. We also conducted semi-structured
interviews to gain an understanding of students' perceptions of
the intervention. The interviews revealed that students benefited
from the intervention in numerous ways, including improved academic performance, mental health, and development of soft skills.
Students adopted the intervention to get more feedback, satisfy
their curiosity, or use their available time. The main reasons for not
adopting the intervention include having other competing deadlines, the intervention not being worth any marks, and feeling
confident about their work.

\section*{Keywords}
time management, procrastination, optional early feedback, programming assignments, self-regulated learning

\section{Introduction}
Introduction
Academic procrastination is prevalent among undergraduate computer science (CS) students. Several meta-analyses have documented
the negative consequences of procrastination, including poor academic performance [17, 24] and well-being [14, 27, 30]. Procrastination can force students to prioritize achieving good grades with
minimal effort over genuine learning, potentially creating challenges for future academic pursuits and professional careers.

Upper-year undergraduate CS students are especially prone to
the harmful effects of procrastination. Many upper-level CS courses
require students to complete large and complex programming assignments. These assignments provide valuable opportunities for
students to apply the theoretical concepts learned in lectures. However, anecdotal evidence suggests that many students don't start
working on these assignments until a few days before the deadline.
Such procrastination leads to insufficient time to debug and test
their code, a lack of opportunity to seek help from course staff, and
a spike in physical and mental health issues. One may attribute such
procrastination to a failure of time management and self-regulation.

While numerous interventions such as extra-credit opportunities, reflective exercises, email reminders, and varying deadline
structures have been employed to reduce academic procrastination
in computer science courses, their effectiveness has been mixed
and often context-dependent. In particular, previous work has either tied early deadlines to grading incentives or provided limited
feedback, leaving unanswered questions about the role of flexible,
non-graded early feedback in shaping students' time management
and learning outcomes. Moreover, there is a lack of targeted research on procrastination mitigation in advanced undergraduate
programming assignments, where the stakes, complexity, and self-
regulatory demands are heightened. This study addresses these
gaps by investigating whether optional early feedback deadlines,
unlinked to grades but offering substantive automated feedback, can
motivate timely engagement and improve academic performance
among upper-year computer science students [2, 20].

We designed an intervention to combat academic procrastination on programming assignments. We applied the intervention to
an upper-level Introduction to Artificial Intelligence course with
four substantial programming assignments, each available for three
weeks. Drawing inspiration from prior work [7, 26], we designed
our intervention to consist of two optional early feedback deadlines. These early deadlines were not worth any marks. However, if
students submitted their work before an early deadline, they would
receive extra feedback in the format of unit test results. These unit
tests evaluated the quality of the submission and helped students
identify mistakes in their submissions.

The early feedback deadlines can mitigate procrastination in
several ways. First, these early deadlines divided each assignment
into smaller, manageable chunks, potentially making it easier for
students to get started and reducing students' stress. Second, early
feedback was exclusively available during that particular week,
motivating students to start their assignments early. Third, detailed
test results in early feedback could help students correct mistakes
in their programs and improve their grades on assignments.

We applied this intervention in the Fall (Sep-Dec) 2021 offering
of the course and evaluated the effects of this intervention using
the Spring (May-Aug) 2021 offering of the course as a control group.
Our work addresses the following research questions.

RQ1: Do the early feedback deadlines motivate students to start
their assignments early?

RQ2: Do the early feedback deadlines improve students' assignment
grades?

\section{Related Work}
In this section, we review the literature relevant to our study, focusing on three key areas: academic procrastination, interventions
designed to mitigate procrastination and self-regulated learning
theory. These topics provide the foundation for understanding the
mechanisms and potential impacts of our proposed early feedback
intervention.

\subsection{Academic Procrastination}
According to Steel [28], academic procrastination refers to voluntarily delaying an intended course of study-related action despite expecting to be worse off for the delay. Academic procrastination negatively affects students' academic performance and well-being [28].
Some negative consequences include lower grades, health issues
(e.g. stress and sleep problems), affective consequences (e.g. anxiety and anger), and problems in their private lives (e.g. lack of
social networks, negative reactions from others) [13, 16–19, 25].
Unfortunately, procrastination is widespread among undergraduate students, affecting more than 70\% of them [22, 25].

Numerous factors cause procrastination, such as a lack of interest in the task [27], the task's urgency or deadline proximity [27],
depression, fatigue, and low self-esteem [10, 27], perfectionism [12],
locus of control [5, 11], goal management abilities [15], neuroticism, openness to experiences, agreeableness, extraversion, conscientiousness [28], lack of academic self-efficacy and poor self-regulation skills [18], fear of failure, task aversiveness or laziness,
lack of assertiveness, resentment towards control from others, and a
desire to take risks [23], and excessive use of social networking sites
for interactive information seeking, discussion, and sharing [29].

\subsection{Procrastination Interventions}
Prior research has studied many creative interventions to combat
procrastination in CS courses, including extra credits [1], reflective
writing assignments and schedule sheets [20], email reminders
[20, 31], and early deadlines with or without feedback [7, 26].

Several interventions did not have significant effects on students'
behaviour. For instance, Allevato and Edwards [1] showed that extra
credits had no significant effects on when students started or completed the programming assignments in an introductory CS course.
Similarly, Martin et al. [20] demonstrated that reflective writing
assignments and schedule sheets did not significantly impact students' behaviour on programming assignments in a junior-level
data structures course.

Email reminders led to some positive results. On the one hand,
email reminders resulted in a significant increase in the number of
early assignment submissions and a significant decrease in the number of late assignment submissions in a junior-level data structures
course [20]. On the other hand, email reminders did not encourage
early submissions while increasing homework attempts in a CS1
course [31].

Our work is most similar to that of Denny et al. [7] and Shaffer and Kazerouni [26], who studied the effects of setting early
deadlines on programming assignments.

Denny et al. [7] created optional early deadlines for the final
programming assignment in an introductory CS course. The early
deadlines were not worth any marks, but the number of passing
unit tests was revealed through automated feedback. The results
showed that the optional early deadlines encouraged many students
to submit their assignments early. Moreover, students who submitted their assignments early were more likely to achieve higher
performance on the assignments.

Shaffer and Kazerouni [26] implemented early deadlines for programming projects in a third-year data structures course. The early
deadline was worth a small percentage of the final project grade (at
most 10\%). The early deadlines resulted in significantly reduced late
submission rates, higher project grades, and higher course grades
for high-performing students.

\subsection{Self-Regulated Learning}
Self-regulated learning (SRL) theory focuses on how learners actively manage their cognitive, emotional, and behavioural processes
to achieve academic goals. Zimmerman [32] defines SRL as the self-
directive process by which learners transform mental abilities into
academic skills, involving goal setting, progress monitoring, and
strategy adjustment. SRL includes three phases: forethought, performance, and self-reflection. Cognitive strategies like rehearsal,
elaboration, and organization aid knowledge acquisition, while
meta-cognitive strategies involve planning, monitoring, and regulating activities [21]. Motivational factors such as self-efficacy, goal
orientation, and behaviours like time management and help-seeking
are crucial [33].

SRL theories trace back to the social cognitive theory by Bandura et al. [4], emphasizing observational learning, and Vygotsky's
sociocultural theory, highlighting the social context of learning.
Contemporary models like Zimmerman's and Pintrich's integrate
motivation with cognitive and metacognitive strategies, enriching
SRL theories.

Empirical studies show SRL's effectiveness in academic settings.
Students using SRL strategies perform better academically [8]. Meta-
analyses, like that of Dignath et al. [9], reinforce the positive impact
of SRL interventions. Programs that enhance SRL skills, including
strategy instruction, modelling, and scaffolded practice [34], significantly improve self-efficacy and performance [6]. Technology
integration, such as e-learning platforms, supports SRL by providing immediate feedback and personalized learning paths [3].

\section{Research Methods}
We studied a fourth-year Introduction to Artificial Intelligence
course at a large research-intensive university in North America.
This course's students primarily consist of CS students in their
third and fourth years of studies and a small number of advanced
engineering students.

We analyzed data from two offerings of this course. The control
group (without the early feedback deadlines) was from the Spring
(May-Aug) 2021 semester and had 239 students. The intervention
group (with early feedback deadlines) was from the Fall (Sep-Dec)
2021 semester and had 145 students. The first author was the sole
instructor for both course offerings. Both offerings were delivered
asynchronously online during the Covid-19 pandemic.

\subsection{Assignment Design}
This course had four substantial programming assignments in
Python, each worth 10\% of a student's final grade. The assignment
topics are described below.

\begin{itemize}
\item A1: solve the rush hour sliding puzzle using search.
\item A2: classify a yeast (control) or astronomy (intervention)
data set using a decision tree.
\item A3: locate a robot by performing inference in a Bayesian
network.
\item A4: explore the Wumpus grid world using reinforcement
learning.
\end{itemize}

Since the two semesters were consecutive, we changed some assignment topics to ensure the originality of the submitted work. For A2,
the two data sets are highly similar since most of the features are
real-valued. A3 in the control semester required students to implement the forward-backward algorithm for hidden Markov models.
In contrast, A3 in the intervention semester required the implementation of the variable elimination algorithm for any Bayesian
network.

Each assignment was available for three weeks and required students to implement a complex algorithm from scratch. We designed
the assignments to help students get started and make progress. We
will describe the design decisions using A1 on search as an example.
Assignment 1 asked students to implement a program to solve a
rush hour sliding puzzle using depth-first search and A* search.
First, we provided a PDF file consisting of detailed descriptions of
the rush hour sliding puzzle. This file described the components of
a rush hour puzzle and the input file to the program. Second, we
provided a “board.py” file, which includes classes for the rush hour
puzzle and functions to read in a puzzle from a file. We decided
to provide these codes so that students can focus on implementing the search algorithm. Furthermore, we provided another file
called “solve.py”, which students will complete. This file contains
several function headers described below. Each function header
includes detailed descriptions of the function's expected behaviour,
the parameters' types and meanings, and the return value's type
and meaning.

\begin{enumerate}
\item is\_goal(state) returns true if and only if the given state
is a goal.
\item get\_path(state) returns a sequence of states from the initial state to the given state.
\item blocking\_heuristic(state) returns the value of the blocking heuristic for the given state.
\item get\_successors(state) returns the successors of the given
state.
\item  dfs(init\_board) executes a depth-first search given an
initial board and returns a sequence of states from the initial
state to a goal state.
\item a\_stari(init\_board, hfn) executes A* search given an
initial board and a heuristic function and returns a sequence
of states from the initial state to a goal state.
\end{enumerate}

The function headers provide scaffolding so that students can tackle
the assignment step by step, in the order of increasing difficulty. In
particular, they can use the earlier functions as helpers for the later
functions.

Providing the function headers allows us to evaluate students'
submissions with automated unit tests. We can also award partial
marks even if a student's submission is incomplete. In the control
semester, we evaluated each student submission with two types
of unit tests: public and hidden. The public tests provided simple
sanity checks (e.g. the submission executed and produced outputs
in the required formats). The hidden tests verified the submission's
correctness and performance in realistic scenarios. For each function, we created one public test and several hidden tests depending
on the difficulty of implementing the function. We reserve most of
the tests to be hidden to encourage the students to write their own
unit tests. The public tests were available throughout the three-
week period, whereas the hidden tests were only used to generate
the assignment grades after the final deadline. In the provided PDF
file, we described the number of public and hidden tests for each
function and the number of marks for each test.

\subsection{The Early Feedback Deadlines}
Although the function headers are designed to help students make
progress, students could still wait until the last few days to start
working on them. To encourage students to start the assignments
early, we introduced two early deadlines for each assignment in
the intervention semester. The two early deadlines are at the end of
the first and second weeks, respectively. Our goal is to encourage
students to complete part of the work during the first two weeks
instead of leaving all the work to the last week.

We divided the provided function headers into two sets to implement the early feedback deadlines. The functions in the first
set were relatively easy to implement. The functions in the second
set were more complex to implement and built on the first set of
functions. Using A1 on search as an example, the first set consisted
of the first four functions.
\begin{enumerate}
\item is\_goal(state)
\item  get\_path(state)
\item blocking\_heuristic(state)
\item get\_successors(state)
\end{enumerate}
The second set consisted of the last two functions implementing
the search algorithms.
\begin{enumerate}
\item dfs(init\_board)
\item  a\_stari(init\_board, hfn)
\end{enumerate}
We then created three projects for each assignment on Marmoset
(our online auto-grading system), corresponding to the two early
feedback deadlines and the final assignment deadline. The first
project was available during the first week only and provided additional public tests for the first set of functions. Similarly, the second
project was available during the second week only and provided
additional public tests for the second set of functions. Note that the
additional public tests in the first two projects differ from the public
tests we use to generate the student's assignment grade in the third
project. The third project was available for the entire three-week
period. Students must submit this project to receive a grade for the
assignment. Before the final assignment deadline, students could
submit the third project to see the results of the public tests. After
the final assignment deadline, students could view the results of
the hidden tests and their assignment grades. The design encourages students to tackle the assignment progressively over the three
weeks. Ideally, students would focus on completing and testing the
easy functions during the first week, followed by completing and
testing the difficult functions during the second week. They could
dedicate the third week to finalizing and testing the entire program.

We made several deliberate choices in the intervention design,
drawing inspiration from prior work [7, 26]. First, our early deadlines provided detailed results of the unit tests, in contrast to prior
work that only provided the number of passing tests [7]. While minimal feedback may be sufficient to motivate novice programmers to
make progress, we believe that detailed feedback is crucial to help
upper-year CS students identify unique errors and improve their
work. Second, the early feedback deadlines were not worth any
marks, whereas prior work made them worth a small percentage
of the assignment grade [26]. Making the early deadlines optional
avoids increasing students' stress and allows students to miss the
early deadlines without losing marks. This flexibility is crucial for
students with a high course load and/or obligations outside of university. Finally, the results of the additional public tests for each
early deadline were exclusively available during their respective
week. For instance, if students did not submit any code during
the first week, they would not have access to the public test results for the first early deadline. The design encourages students to
start the assignments early so they don't miss the early feedback
opportunity.

Overall, we made these design decisions to achieve a delicate
balance between offering the flexibility of not meeting the early
deadlines and providing sufficient incentives for students to make
early submissions. Our intervention can be seen as a compromise
between that of Denny et al. [7] and Shaffer and Kazerouni [26]. In
contrast to Shaffer and Kazerouni [26], we provided flexibility by
making the early deadlines not worth any marks. Unlike Denny
et al. [7], we provided detailed test results if students made an early
submission during the first two weeks.

\subsection{Semi-Structured Interviews}
We conducted semi-structured interviews with a small group of students to explore their perceptions of procrastination, programming
assignments, and the intervention. Two undergraduate research
assistants carried out the interviews to create an open environment
and encourage honest discussions about students' views on the
course and their tendencies to procrastinate. This approach was
chosen because students might feel hesitant to share their opinions
directly with faculty. The research assistants conducted the interviews remotely via Microsoft Teams during the summer of 2022.
Each interview lasted between 30 minutes and an hour, allowing for
in-depth conversations and a thorough exploration of different aspects of each student's experiences. Twenty-one students from the
control group and twelve from the intervention group participated
in the interviews.

We analyzed the data using grounded theory qualitative analysis.
Three coders—one author of this paper and two undergraduate
research assistants—independently coded all the interview transcripts. Afterwards, they discussed and adjusted the results until
reaching a consensus. In this paper, we present the analysis of the
interviews with students in the intervention group. Specifically, we
discuss students' perceptions of the positive and negative effects of
the early feedback deadlines and why students chose to use or not
use each deadline.
Coding procedures for interview transcripts followed grounded
theory guidelines, with three independent coders reviewing transcripts before resolving discrepancies and reaching consensus. Intercoder agreement was not formally measured, but consensus discussions ensured coding validity.

\section{Results and Discussion}
In this section, we address RQ1 by analyzing when students started
working on assignments in Section 4.1. We address RQ2 by comparing and analyzing grades between control and intervention groups,
as well as within the intervention group in section 4.2.

\subsection{Starting Assignments Early}
To address RQ1, we analyzed the effects of the early feedback deadlines on when students started working on the assignments. We
used the date of a student's first submission on Marmoset as an
approximate measure of when they started working on each assignment. Table 1 summarizes the percentage of students who started
each assignment each week. There is a small percentage of students
who did not submit each assignment. We included this percentage
as a last category titled “Did not submit.” The two values in each
cell are for the control and intervention groups, respectively.

Comparing the two semesters, there is a stark contrast in the
percentage of students starting each assignment early. Less than
8\% of the students in the control group started each assignment
during the first week, whereas more than 14\% of the students in the
intervention group started each assignment during the first week.
The intervention increased the percentage of students who started
each assignment during the first two weeks from 13-23\% in the
control group to 44-53\% in the intervention group.

We hypothesize that the intervention had a significant effect
on motivating students to make the first submission early. To
test this hypothesis, we used Pearson's chi-squared test to determine, for each assignment, if there was a statistically significant difference in the percentages of students who made their
first submission each week (or who didn't submit each assignment) between the control and the intervention groups. The test
revealed that there was a statistically significant difference in the
percentage of students making their first submissions during each
week between the control and the intervention groups for all four
assignments (DoF = 3, $N = 384$, A1: $\chi^2 = 72.74$, $p < 0.001$, A2: $\chi^2 = 164.66$, $p < 0.001$, A3: $\chi^2 = 1976.81$, $p < 0.001$, A4: $\chi^2 = 85.72$, $p < 0.001$).

\subsection{Improving Assignment Grades}
To address RQ2, we compared the assignment grade distributions
between different groups of students.
\subsubsection{Comparing Grades Between Control and Intervention Groups.}
First, we compared the assignment grades between the two semesters.

\begin{table}[h!]
\centering
\begin{tabular}{lcccc}
\hline
 & A1 & A2 & A3 & A4 \\
\hline
\textbf{Control} & & & & \\
Started in Week 1 & 8\% & 3\% & 0\% & 6\% \\
Started in Week 2 & 15\% & 10\% & 17\% & 10\% \\
Started in Week 3 & 75\% & 84\% & 76\% & 78\% \\
Did not submit     & 2\% & 3\% & 7\% & 6\% \\
\hline
\textbf{Intervention} & & & & \\
Started in Week 1 & 30\% & 22\% & 14\% & 18\% \\
Started in Week 2 & 14\% & 27\% & 39\% & 30\% \\
Started in Week 3 & 51\% & 48\% & 41\% & 41\% \\
Did not submit     & 5\% & 3\% & 6\% & 11\% \\
\hline
\end{tabular}
\caption{Percentage of students who started each assignment each week or did not submit the assignment.The top portion
shows the percentages for the control semester, whereas the
bottom portion shows the percentages for the intervention
semester.}
\label{tab:assignment_start}
\end{table}

See Table 2 for the median assignment grades for both semesters.
Running the Shapiro-Wilk test revealed that the grade distributions
were not normal ($p < 0.001$). Therefore, we used the Wilcoxon
rank-sum test to compare the grade distributions for each assign-
ment between the two semesters. The test revealed that there were
no significant differences in grade distributions between the two
semesters for all the assignments except A3 (A1: $p = 0.750$, A2: $p = 0.948$, A3: $p = 0.027$, A4: $p = 0.809$).

\begin{table}[h!]
\centering
\begin{tabular}{|c|c|c|}
\hline
\textbf{Assignment} & \textbf{Control} & \textbf{Intervention} \\
\hline
A1 & 98\% & 98\% \\
A2 & 85\% & 91\% \\
A3 & 91\% & 97\% \\
A4 & 100\% & 100\% \\
\hline
\end{tabular}
\vspace{0.5cm}
\caption{Median Assignment Grades For Both Semesters}
\label{tab:median_grades}
\end{table}

One reason for the lack of difference between the grade distributions may be the ceiling effect. If students are already achieving
high grades in the control semester, there is little room for intervention to improve their grades in the intervention semester. As
shown in Table 2, the median grades for all assignments in the
control semester were quite high, with A1 and A3 median grades
very close to 100\%. With such high grades in the control semester,
it is challenging to discern the true impact of the intervention on
the students' performance.

Our result contrasts that of Shaffer and Kazerouni [26], who
reported a significant difference in the grade distributions between
the control and intervention groups. One key difference between
the two studies is that our early deadlines are not worth marks,
whereas Shaffer and Kazerouni [26] made their early deadlines
worth a small percentage of the final project grades. Contrasting the
two results, we conjecture that having mandatory early deadlines
may be the key driving factor in improving students' assignment
grades significantly. However, the trade-off is that mandatory early
deadlines could negatively impact students' mental health. We leave
the exploration of this trade-off to future work.

Another difference between our work and that of Shaffer and
Kazerouni [26] is that we studied a fourth-year CS course, whereas
they studied a third-year CS course. Fourth-year students typically
have more work experience and advanced coursework than their
lower-year counterparts. This additional experience can influence
their time management skills, familiarity with complex assignments,
and overall academic performance, potentially skewing the results.
Consequently, the intervention's effects might appear differently
across different year levels, with fourth-year students potentially
benefiting less due to their already-developed skills and strategies.
Future studies should consider stratifying the analysis by year level
to understand better how the intervention impacts students at
different stages of their academic careers.

\subsubsection{Comparing Grades Within the Intervention Group. }
Next, we focused on the intervention group and investigated whether meeting at least one early feedback deadline improved students' assignment grades significantly. After dividing the students into groups
based on the week they made their first submission, we compared
the grade distributions among the groups using the Wilcoxon Rank-
Sum Test. Table 3 shows the median grades for each assignment,
given that students made their first submission during a particular
week.
\begin{table}[h!]
\centering
\begin{tabular}{|c|c|c|c|}
\hline
\textbf{Assignment} & \textbf{Week 1} & \textbf{Week 2} & \textbf{Week 3} \\
\hline
A1 & 98\% & 100\% & 93\% \\
A2 & 95\% & 95\% & 73\% \\
A3 & 97\% & 97\% & 97\% \\
A4 & 100\% & 100\% & 97\% \\
\hline
\end{tabular}
\vspace{0.5cm}
\caption{Median Assignment Grades For Each Assignment and Students Starting Assignments Each Week in the Intervention Group}
\label{tab:intervention_medians}
\end{table}

Our results showed that students who used at least one early
feedback deadline achieved significantly higher grades than those
who didn't use any early deadline for several assignments. In particular, students who started in week 1 achieved significantly higher
grades than those who started in week 3 for A1, A2, and A3 (A1: $p = 0.031$, A2: $p = 0.007$, A3: $p = 0.031$, A4: $p = 0.078$). Similarly,
students who started in week 2 achieved significantly higher grades
than those who started in week 3 for A2 and A4 (A1: $p = 0.096$, A2: $p = 0.042$, A3: $p = 0.285$, A4: $p = 0.003$).

Next, we did a similar analysis by combining students who used
at least one early deadline into one group. The results revealed that
students who used at least one early deadline obtained significantly
higher grades than those who did not for A1, A2, and A4 (A1: $p = 0.015$, A2: $p = 0.004$, A3: $p = 0.088$, A4: $p = 0.003$).

So far, our results suggest that using at least one early deadline
improved the students' assignment grades significantly within the
intervention group. These results are similar to that of Denny et al.
[7], who found that the early feedback on the number of passing
tests significantly improved students' assignment scores. However,
we did not control when a student started each assignment. As a
result, when interpreting these results, we cannot rule out the possibility that stronger students self-selected to start the assignments
early and, as a result, achieved significantly higher grades.

We performed a within-student analysis to understand further
what caused the students who started early to achieve higher grades.
For each pair of assignments, we isolated students who used the intervention for one assignment and did not use the intervention for
the other assignment. We normalized all the grades by computing
their z-scores. Then, we compared the students' grade distributions for the two assignments using the Wilcoxon Rank-Sum Test.
If the intervention caused students to achieve higher grades, we
would expect that the students received significantly higher grades
on the assignment where they used the intervention than on the
assignment where they did not.

Unfortunately, we did not observe a significant difference in the
grade distributions for any pair of assignments. Therefore, it is possible that our intervention primarily motivated stronger students to
start the assignments earlier than they usually would, thus achieving higher grades than those who did not start the assignment early.
One important future direction is to design and assess interventions
that could motivate weaker students to start the assignments early
and improve their learning.

\subsubsection{Which Factor Was More Important: Starting Early or Getting
Additional Feedback?}
Our intervention provided two benefits to
the students. First, it motivates students to start the assignment
early, thus leaving more time for testing, debugging and seeking
help. Second, the intervention offers additional feedback that could
help students verify their solution quality and find problems in
their solutions. Therefore, one might wonder which of the two
factors (starting early or getting additional feedback) was more
critical in improving students' grades. To untangle the effects of
these two factors, we performed a grade comparison between the
students in the control group, similar to the analysis in section
4.2.2. We divided the students based on the week they started each
assignment and compared the grade distributions among the groups
using the Wilcoxon Rank-Sum Test. If students who started the
assignments early in the control semester achieved significantly
higher grades than those who did not, then we can infer that the
additional feedback provided in the intervention semester may not
be critical in improving students' assignment grades.

We found no significant difference in the grade distributions
between the students who started in week 1 and week 3 (A1: $p = 0.236$, A2: $p = 0.621$, A3: N/A, A4: $p = 0.163$). (We could not perform
the analysis for A3 since no student started A3 in week 1 in the
control semester.) Similarly, there was no significant difference in
the grade distributions between the students who started in week
2 and week 3 (A1: $p = 0.499$, A2: $p = 0.270$, A3: $p = 0.159$, A4: $p = 0.060$). Moreover, we compared students who used at least one
early feedback deadline with the rest. Between these two groups,
there was a significant difference in the grade distributions for A4
only (A1: $p = 0.242$, A2: $p = 0.236$, A3: $p = 0.159$, A4: $p = 0.024$).

A comparison of the two analyses for the control and intervention groups suggests that additional feedback was essential in
improving the grades of the students who started early compared
to those who did not. Based on the analysis of the control group,
we can conclude that starting the assignments earlier alone did
not significantly improve students' assignment grades. As for the
intervention group, we cannot fully untangle the effects of receiving additional feedback and starting early since the opportunity to
receive additional feedback motivated more students to start the assignments early. However, we could conclude that the combination
of starting early and getting more feedback improved the students'
grades significantly relative to the rest of the students.

\section{Limitations}
The differences in the course offerings could significantly impact
students' behaviour and performance in both groups. Both courses
were taught by the same instructor, which helped minimize the
difference in the course materials and delivery style. Although we
tried to minimize the difference in the assignment topics, we had
to make some changes to ensure the originality of work between
the two consecutive semesters.

A student's first submission time might not be a great proxy for
when they started the assignment. The student could have worked
on the assignment offline for a period of time before making the
first submission on Marmoset. Conversely, some students may have
made an empty first submission on Marmoset to test the system
before they start working on the assignment. However, the first
submission time was the best data to estimate when a student
started working on the assignment. Future research should consider
triangulating assignment start measures by collecting direct student
self-reports or leveraging digital trace data (e.g., version control
commit history), which would improve the accuracy of time-on-task
and engagement analyses.

Due to the COVID-19 pandemic, this course was delivered asynchronously online for both semesters. The delivery mode may have
impacted students in several ways. Some students may have struggled with the programming assignments more than usual and had
trouble seeking help. Some students may have sought unauthorized help from their friends or other outside sources. Additionally,
the unique circumstances surrounding the COVID-19 pandemic,
including changes in motivation, stress levels, and the virtual learning environment, may have influenced both engagement with the
intervention and overall procrastination behaviors. These factors
limit the generalizability of our findings to post-pandemic settings
or traditional in-person course structures.

It is challenging to argue whether our results can be generalized
to other settings. However, our study benefits from having a large
sample size, which increases the reliability of our findings and helps
address the threat to external validity.

Our study is a quasi-experiment since the students were not
randomly assigned to the control and intervention groups. However,
our experimental design and large sample size give us reasonable
confidence in establishing a causal relationship between the early
feedback deadlines and the observed positive impacts on students'
assignment performance.

We further acknowledge that the interview analysis presented
here focused mainly on students in the intervention group with a
limited sample size, potentially excluding diverse perspectives, especially
from students less motivated to participate in early interventions
or qualitative research. Expanding interview participation and including more control group voices would enhance the validity and
richness of qualitative findings.

\section{Conclusion and Future Work}
In this work, we designed and evaluated an intervention to combat procrastination in a fourth-year CS course. Our intervention
provided optional early feedback if students submitted their work
early. We conducted a quasi-experiment to evaluate the effects of
the intervention by comparing students' behaviour between the
control and intervention groups. We found that the intervention
significantly motivated more students in the intervention group
to start the assignments early than the control group. Unfortunately, the intervention did not result in a significant difference in
the students' grades between the control and intervention groups.
However, within the intervention group, students who met at least
one early deadline achieved significantly higher assignment grades
than those who did not meet any early deadlines. We performed further analyses to disentangle the effects of starting the assignments
early and receiving additional feedback. Our study showed that
receiving additional feedback is crucial for significantly improving
students' grades. While our results highlight the immediate benefits
of optional early feedback for motivating timely engagement, they
also suggest a need for further exploration of how such interventions scale to other contexts and student populations. Specifically,
future replications in in-person or blended learning environments
will be necessary to test the robustness and external validity of
these findings beyond the pandemic-era, asynchronous delivery
model.

We interviewed some students to understand their perceptions of
the intervention. Our results showed that students benefited from
the intervention in several ways, including improved academic
performance, mental health, and soft skills. Students experienced
a few negative effects of the intervention, most indicating a slight
decline in their mental health. Next, we asked students why they
used or did not use the intervention. On the one hand, students
used the intervention to get more feedback, satisfy their curiosity,
or use their available time. On the other hand, the main reasons
for not adopting the intervention include having other competing
deadlines, the intervention not being worth any marks, and feeling
confident about their work.

Fourth-year students typically have more work experience and
advanced coursework under their belts compared to their lower-year counterparts. This additional experience can influence their
time management skills, familiarity with complex assignments,
and overall academic performance, potentially skewing the results.
Consequently, the intervention's effects may appear differently
across different year levels, with fourth-year students potentially
benefiting less due to their already developed skills and strategies.
Future studies should consider stratifying the analysis by year level
to better understand how the intervention impacts students at
different stages of their academic careers.

We are excited to explore several future directions. First, one
future direction is to study the trade-off between setting mandatory
and optional early deadlines. We aim to design interventions to optimize the trade-off between the impacts on academic performance
and mental health. Moreover, we would like to understand how our
intervention affects the strong and weak students differently and
to design interventions to improve the performance of the weaker
students. Another pressing question for future research is how
to design targeted support mechanisms for students who are less
likely to self-select into early action, such as lower-performing, less
confident, or less engaged students. Integrating adaptive scaffolding, personalized reminders, or peer-support elements may help
address this challenge. Finally, another future direction is to design interventions to teach students effective strategies to manage
competing deadlines.

\section*{Acknowledgments}
We thank Michelle Craig and Jonathan Calver for helpful feedback
on earlier drafts of this paper.


\begin{thebibliography}{99}
\bibitem{ref1} Anthony Allevato and Stephen H Edwards. 2013. The effects of extra credit opportunities on student procrastination. In 2013 IEEE Frontiers in Education Conference (FIE). IEEE, IEEE, Oklahoma City, OK, 1831--1836.
\bibitem{ref2} Arpana Amarnath, Sevin Ozmen, Sascha Y Struijs, Leonore de Wit, and Pim Cuijpers. 2023. Effectiveness of a guided internet-based intervention for procrastination among university students--A randomized controlled trial study protocol. Internet Interventions 32 (2023), 100612.
\bibitem{ref3} Roger Azevedo, Daniel C Moos, Amy M Johnson, and Amber D Chauncey. 2010. Measuring cognitive and metacognitive regulatory processes during hypermedia learning: Issues and challenges. Educational psychologist 45, 4 (2010), 210--223.
\bibitem{ref4} Albert Bandura et al. 1986. Social foundations of thought and action. Englewood Cliffs, NJ 1986, 23--28 (1986), 2.
\bibitem{ref5} Randy Carden, Courtney Bryant, and Rebekah Moss. 2004. Locus of control, test anxiety, academic procrastination, and achievement among college students. Psychological Reports 95, 2 (2004), 581--582.
\bibitem{ref6} Timothy J Cleary and Peter Platten. 2013. Examining the correspondence between self-regulated learning and academic achievement: A case study analysis. Education Research International 2013, 1 (2013), 272560.
\bibitem{ref7} Paul Denny, Jacqueline Whalley, and Juho Leinonen. 2021. Promoting early engagement with programming assignments using scheduled automated feedback. In Proceedings of the 23rd Australasian Computing Education Conference. ACM, Australia, 88--95.
\bibitem{ref8} Amy L Dent and Alison C Koenka. 2016. The relation between self-regulated learning and academic achievement across childhood and adolescence: A meta-analysis. Educational Psychology Review 28 (2016), 425--474.
\bibitem{ref9} Charlotte Dignath, Gerhard Buettner, and Hans-Peter Langfeldt. 2008. How can primary school students learn self-regulated learning strategies most effectively?: A meta-analysis on self-regulation training programmes. Educational Research Review 3, 2 (2008), 101--129.
\bibitem{ref10} Joseph R Ferrari. 2000. Procrastination and attention: Factor analysis of attention deficit, boredomness, intelligence, self-esteem, and task delay frequencies. Journal of Social Behavior and Personality 15, 5 (2000), 185.
\bibitem{ref11} Joseph R Ferrari, James T Parker, and Carolyn B Ware. 1992. Academic procrastination: Personality correlates with Myers-Briggs types, self-efficacy, and academic locus of control. Journal of Social Behavior and Personality 7, 3 (1992), 495--502.
\bibitem{ref12} Gordon L Flett and Paul L Hewitt. 2002. Perfectionism and maladjustment: An overview of theoretical, definitional, and treatment issues. (2002).
\bibitem{ref13} Gordon L Flett, Murray Stainton, Paul L Hewitt, Simon B Sherry, and Clarry Lay. 2012. Procrastination automatic thoughts as a personality construct: An analysis of the procrastinatory cognitions inventory. Journal of Rational-Emotive \& Cognitive-Behavior Therapy 30 (2012), 223--236.
\bibitem{ref14} Carola Grunschel, Justine Patrzek, and Stefan Fries. 2013. Exploring reasons and consequences of academic procrastination: An interview study. European Journal of Psychology of Education 28 (2013), 841--861.
\bibitem{ref15} Daniel E Gustavson and Akira Miyake. 2017. Academic procrastination and goal accomplishment: A combined experimental and individual differences investigation. Learning and Individual Differences 54 (2017), 160--172.
\bibitem{ref16} Eunkyung Kim and Eun Hee Seo. 2013. The relationship of flow and self-regulated learning to active procrastination. Social Behavior and Personality: an International Journal 41, 7 (2013), 1099--1113.
\bibitem{ref17} Kyung Ryung Kim and Eun Hee Seo. 2015. The relationship between procrastination and academic performance: A meta-analysis. Personality and Individual Differences 82 (2015), 26--33.
\bibitem{ref18} Robert M Klassen, Lindsey L Krawchuk, and Sukaina Rajani. 2008. Academic procrastination of undergraduates: Low self-efficacy to self-regulate predicts higher levels of procrastination. Contemporary Educational Psychology 33, 4 (2008), 915--931.
\bibitem{ref19} Katrin B Klingsieck, Axel Grund, Sebastian Schmid, and Stefan Fries. 2013. Why students procrastinate: A qualitative approach. Journal of College Student Development 54, 4 (2013), 397--412.
\bibitem{ref20} Joshua Martin, Stephen H Edwards, and Clfford A Shaffer. 2015. The effects of procrastination interventions on programming project success. In Proceedings of the eleventh annual International Conference on International Computing Education Research. 3--11.
\bibitem{ref21} Paul R Pintrich. 2000. Multiple goals, multiple pathways: The role of goal orientation in learning and achievement. Journal of educational psychology 92, 3 (2000), 544.
\bibitem{ref22} Laura A Rabin, Joshua Fogel, and Katherine E Nutter-Upham. 2011. Academic procrastination in college students: The role of self-reported executive function. Journal of Clinical and Experimental Neuropsychology 33, 3 (2011), 344--357.
\bibitem{ref23} Sonia Rahimi and Nathan C Hall. 2021. Why are you waiting? Procrastination on academic tasks among undergraduate and graduate students. Innovative Higher Education 46 (2021), 759--776.
\bibitem{ref24} Michelle Richardson, Charles Abraham, and Rod Bond. 2012. Psychological correlates of university students' academic performance: a systematic review and meta-analysis. Psychological bulletin 138, 2 (2012), 353.
\bibitem{ref25} Gregory Schraw, Theresa Wadkins, and Lori Olafson. 2007. Doing the things we do: A grounded theory of academic procrastination. Journal of Educational Psychology 99, 1 (2007), 12.
\bibitem{ref26} Clifford A Shaffer and Ayaan M Kazerouni. 2021. The impact of programming project milestones on procrastination, project outcomes, and course outcomes: A quasi-experimental study in a third-year data structures course. In Proceedings of the 52nd ACM Technical Symposium on Computer Science Education. ACM, Virtual, 907--913.
\bibitem{ref27} Piers Steel. 2007. The nature of procrastination: a meta-analytic and theoretical review of quintessential self-regulatory failure. Psychological Bulletin 133, 1 (2007), 65.
\bibitem{ref28} Piers Steel and Katrin B Klingsieck. 2016. Academic procrastination: Psychological antecedents revisited. Australian Psychologist 51, 1 (2016), 36--46.
\bibitem{ref29} Arminda Suárez-Perdomo, Zuleica Ruiz-Alfonso, and Yaritza Garcés-Delgado. 2022. Profiles of undergraduates' networks addiction: Difference in academic procrastination and performance. Computers \& Education 181 (2022), 104459.
\bibitem{ref30} Dianne M Tice and Roy F Baumeister. 2018. Longitudinal study of procrastination, performance, stress, and health: The costs and benefits of dawdling. In Self-Regulation and Self-Control. Routledge, 299--309.
\bibitem{ref31} Angela Zavaleta Bernuy, Qi Yin Zheng, Hammad Shaikh, Andrew Petersen, and Joseph Jay Williams. 2021. Investigating the Impact of Online Homework Reminders Using Randomized A/B Comparisons. In Proceedings of the 52nd ACM Technical Symposium on Computer Science Education. ACM, Virtual, 921--927.
\bibitem{ref32} Barry J Zimmerman. 2002. Becoming a self-regulated learner: An overview. Theory into practice 41, 2 (2002), 64--70.
\bibitem{ref33} Barry J Zimmerman and Dale H Schunk. 2011. Self-regulated learning and performance: An introduction and an overview. Handbook of self-regulation of learning and performance (2011), 15--26.
\bibitem{ref34} Barry J Zimmerman and Dale H Schunk. 2012. Motivation: An essential dimension of self-regulated learning. In Motivation and self-regulated learning. Routledge, 1--30.
\end{thebibliography}
\end{document}